# Construction of accurate machine learning force fields for copper and silicon dioxide

Wenwen Li[1], Yasunobu Ando[1]

[1] *Research Center for Computational Design of Advanced Functional Materials, National Institute of Advanced Industrial Science and Technology, Tsukuba, Ibaraki 305-8568, Japan*

Recently, the machine learning force field has emerged as a powerful atomic simulation approach for its high accuracy and low computational cost. However, its applications in the multi-components materials are relatively less. In this study, the ML force fields are constructed for both elemental material (Cu) and binary material ($SiO_2$). The atomic environments are described by the structural fingerprint that takes the bond angle into account, and then, different ML techniques, including linear regression, neural network and mixture model method, are used to learn the structure-force relationship. We found that the use of angular structural fingerprint and mixture model method significantly improves the accuracy of ML force fields. In addition, we discussed the effective structural fingerprints auto-selection method based on LASSO and the genetic algorithm. The atomic simulations carried out with ML force fields are in excellent agreement with *ab initio* calculations.

## I. INTRODUCTION

Molecular dynamics (MD) is one of the most powerful tools to reveal the atomic-level structure evolution of physical and chemical processes. Usually, the atomic evolution trajectory is propagated by solving the Newton's equation of motion for each particle, where energy and force are calculated using either classical interatomic potential or *ab initio* methods. The *ab initio* MD provides the most reliable and accurate force information; however, it is prohibitively expensive in the large-scale simulation. The classical interatomic potential, such as Lennard-Jones, Morse, embedded atom model (EAM),[1] is in general much faster and scale better than *ab initio* methods. But, they cannot precisely reproduce quantum-mechanical forces and have limited transferability.

In the past decade, molecular dynamics based on date-driven atomic simulation approaches has attracted much attention as a promising way to achieve high computational accuracy and speed.[2] As illustrated in Fig. 1, the data-driven approach means that the potential or the force field are "trained" using the DFT-calculated energies or forces as the reference data. Then the machine-learning (ML) ones can predict the corresponding atomic properties based on the reference datasets. The contemporary methods can be categorized into two groups, ML potential[3–5] and ML force fields[6–13], according to energy or force is fitted, respectively.

Just like the classical interatomic potential, the ML potentials also determine the energy of a given system, and then the atomic force is obtained from their derivative. One of the most widely-used ML potentials is the high-dimensional neural network (NN) potential.[3,14–30] In such method, generalized symmetry functions are used to describe local atomic environment. The NN potentials have been developed for many different materials, such as Si,[3] C,[31] Cu,[17] ZnO,[20] $TiO_2$,[15] $H_2O$ dimers,[25] $Li_3PO_4$,[30] Cu clusters supported on Zn oxide,[22] and Au/Cu nanoparticles with water molecules[14]. In addition, they have been used to simulate the phase transition, atom diffusion and search for equilibrium structures with not only molecular dynamics but also Monte Carlo,[32] nudged elastic band method[30] and metadynamics[33]. Another popular ML potential is Gaussian approximation potential (GAP), which is based on the bispectrum decomposition and Gaussian process regression.[4] The GAP potentials are developed to describe tungsten and its defects,[34] solid and liquid water,[35] amorphous carbon,[36] and lithiation of carbon anode[37]. In addition, many other ML potentials, such as SVM and Coulomb matrix method,[38] have been developed in the last ten years.

The ML force field is proposed more recently, but its progress in encouraging.[6–13,39–46] Different from ML potentials, the force field learns force directly instead of deriving from the derivative of potential energy surface. The ML force field was firstly proposed independently by V. Botu *et.al*[7] and Z. Li *et.al*.[8] They all used the vector structural descriptors as fingerprint of atomic environments and separately learn individual force components with kernel regression and Gaussian process. Lately, A. Glielmo *et. al.* propose a novel scheme, which predict the forces as vector quantities with Gaussian process regression. In addition, they used the many-body kernel to represent the dependence of force on not only interatomic distance but also bond angles. Since much more force components can be obtained from one DFT calculation than energy, the construction of the ML force field is easier than the ML potentials. The ML force fields have been successfully constructed for many elemental materials, such as Al,[6,7,9,11] Si,[10] U,[6] Cu,[13] *etc*. However, its application in multi-components materials have been only preliminarily discussed yet.[8,12] In addition, although the A. Glielmo's work has revealed that the many body terms effectively improved the accuracy of force field, the structural descriptor for the ML force field made by Botu's scheme with consideration of bond angle is still elusive.

Another exciting progress of data-driven atomic simulation method is a novel mixture model method has been proposed by T. M. Palm et al., which has been successfully used in the construction of ML potentials.[44] With the unsupervised Gaussian mixture model, one atomic environment can be automatically assigned to a group with certain probability. Then, the distinct structure-energy model was built for each group. The mixture model significantly improves the accuracy of the potential energy surface fitting because it reduces the diversity of reference structures. The idea of mixture model might also benefit the construction of ML force field, but, such discussion has not been reported.

In addition, the accuracy, speed and reliability of both ML potentials and force fields depend strongly on the choice of descriptors, which used as input for the machine learning method. However, the descriptors have always been determined in an empirical way, until G. Imbalzano et al. recently proposed automatic protocols to select a number of fingerprints for the NN potential and GAP.[47] The automatically selected descriptors show great potential in the simplification of NN potential and GAP. Therefore, development of the same protocol for the ML force fields is necessary.

In this work, we systematically construct the ML force fields for both elemental (Cu) and binary (SiO$_2$) materials. A new structural descriptor that takes bond angle into consideration is proposed, and it significantly improved the accuracy of atomic force. Moreover, we propose an automatic descriptor selection protocol based on the least absolute shrinkage and selection operator (LASSO) and the genetic algorithm(GA). Applying new descriptors and its selection by LASSO and GA, several ML force fields constructed by linear regression, neural network, and linear mixture regression are compared. Finally, the ML force fields are applied for various atomic simulations.

## II. METHOD:

### A. Reference database

The *ab initio* MD is the most convenient and quick way to sample the reference data for ML force field construction, since abundant reference structures and corresponding quantum mechanism properties (*i.e.* energy and atomic force) can be directly obtained from one piece of *ab initio* MD trajectory. However, the MD steps might have high structural correlation, which is harmful for the robustness and completeness of database. To overcome this problem, a two-step process is adopted.

First, several configurations of the system as initial conditions for MD are built, and then fast *ab initio* MD (with low cutoff energy and loose *k*-points mesh) is carried out to produce atomic configurations of Cu and SiO$_2$ (see Fig. 2). We used the following supercells for Cu: face-centered cubic supercells (32~128 atoms), surface (111) supercells (80~144 atoms), surface (100) supercells (64~120 atoms) and amorphous supercells (32~64 atoms). For SiO$_2$, we used quartz supercells (9~72 atoms), cristobalite supercells (12~96 atoms), quartz surface (001) supercells (27~108 atoms), cristobalite surface (001) supercells (36~144 atoms) and amorphous supercells (18~96 atoms). The MD is carried out at the temperature of 300, 600, 900 and 1200 K with *NVT* ensemble.

Then the reference structures are picked up from the fast MD trajectories for the final calculation of atomic forces. To break the structural correlation, we take out the snapshot of MD steps in each 40 fs. Up to 8-point defects (interstitials and vacancies) are inserted into the structures before the precise atomic forces. Forces are computed with Vienna *ab initio* simulation package (VASP)[48,49]. The DFT calculation adopts the projector augmented wave (PAW) method to treat atomic core electrons, while the Perdew-Burke-Ernzerhof functional within the generalized gradient approximation is adopted to describe the electron-electron interactions.[50,51] The cutoff energy and *k*-point mesh are set so that the convergence of the atomic force within 0.02 eV/Å was achieved.

In total, the Cu reference database contains 22240 structures that includes 1220312 atomic environments. The SiO$_2$ database contains 39390 structures that includes 867986 atomic environments. Both reference databases are randomly split into the training set (80%) and testing set (20%). The components of Cu and SiO$_2$ reference database are listed in the supplementary material.

### B. Structural fingerprint

A number of structural descriptors has been proposed to represent the local atomic environment, such as symmetry functions,[52] bispectrum[53] and SOAP[54]. The vector atomic-fingerprint function (noted as "fingerprint" in this work) proposed by Botu and Ramprasad has been proved to be an effective structural descriptor for the prediction of vectorial atomic properties, such as forces, in solid materials.[7,13] With the fingerprint, the atomic environment of $i^{th}$ atom in specific atomic configuration can be represented by in the following formula:

$$V_i^\alpha = \sum_j \frac{r_{ij}^\alpha}{r_{ij}} e^{-\left(\frac{R_{ij}}{\eta}\right)^2} \cdot f_c(r_{ij}) \qquad (1)$$

Here, $r_{ij}$ is the distance between atoms *i* and *j*, while $r^\alpha_{ij}$ is a scalar projection of this distance along $\alpha$ direction. $\eta$ is a parameter that controls the decay rate, and $f_c$ is the cutoff function that gradually reduce the contribution of distant atoms and truncate the interatomic interaction when $r_{ij}$ is larger than the cutoff distance $R_c$. The formula of cutoff function is:

$$f_c = \begin{cases} 0.5 \times \left[\cos\left(\dfrac{\pi r_{ij}}{R_c}\right) + 1\right] & r_{ij} \leq R_c \\ 0 & r_{ij} > R_c \end{cases} \quad (2)$$

The eq. (1) is very similar to the radial symmetry function, the widely used structural descriptor in the construction of high-dimensional neural network potential,[3,52] except that the direction resolved term $r^\alpha_{ij}/r_{ij}$ is added. The direction resolved term allows the value of eq. (1) changes with the projection direction $\alpha$ so that it is applicable to the representation vectorial properties.

Although the eq. (1) has been proved to be very effective in various materials, it ignores the bond angle information, which might be insufficient for the complex covalent materials. In this study, we modified eq. (1) and proposed a new structural fingerprint that takes the bond angle into consideration. The formula of the new structural fingerprints are as follows:

$$V_i^{1,\alpha} = \sum_j \frac{r_{ij}^\alpha}{r_{ij}} e^{-\eta(R_{ij} - R_s)^2} \cdot f_c(r_{ij}), \quad (3)$$

and,

$$V_i^{2,\alpha} = 2^{1-\zeta} \sum_j \sum_k (\vec{r}_{ij} + \vec{r}_{ik})^\alpha \left(1 + \cos(\theta_{ijk} - \theta s)\right)^\zeta \cdot e^{-\eta\left(\frac{r_{ij}+r_{ik}}{2} - R_s\right)^2} \cdot f_c(r_{ij}) \cdot f_c(r_{ik}). \quad (4)$$

Equations (3) and (4) are named as a radial fingerprint and an angular fingerprint respectively. Here, $r_{ij}$ and $r_{ik}$ are the interatomic distance between $i$ and $j$, $i$ and $k$, respectively. $\theta_{ijk}$ is the angle between bond $ij$ and $ik$. $(\mathbf{r}_{ij}+\mathbf{r}_{ik})^\alpha$ is the scalar projection of vector $\mathbf{r}_{ij}+\mathbf{r}_{ik}$ along $\alpha$ direction. Two parameters of radial fingerprint $V_i^{1,\alpha}$, or $\eta$ and $R_s$, are used to control the width of peak and to shift the peak position. Two additional parameters, i.e. $\zeta$ and $\theta_s$, are used in the angular fingerprint $V_i^{2,\alpha}$. Applying a $\theta_s$ parameter allows probing of specific regions of the angular environment in a similar way as is accomplished with $R_s$ in the radial part. Also, $\zeta$ changes the width of the peaks in the angular environment. Just like the eq (1), the eq. (4) is also a modification of the angular symmetry function proposed in Ref. [29]. We also tested two other forms of direction resolved term, which is $(\mathbf{r}_{ij} \times \mathbf{r}_{ik})^\alpha/|\mathbf{r}_{ij} \times \mathbf{r}_{ik}|$ and $(\mathbf{r}_{ij} - \mathbf{r}_{ik})^\alpha/|\mathbf{r}_{ij} - \mathbf{r}_{ik}|$, but the eq. (4) works the best. It might be because that the other two forms of angular fingerprints does not have the symmetry corresponding to the exchange of atom index $j$ and $k$, while the eq. (4) has.

To distinct different atomic configurations, a spectrum of $V_i^{1,\alpha}$ and $V_i^{2,\alpha}$ fingerprints with different parameters must be used to represent the local environment of an atom. For the elemental system like Cu, the fingerprint set simply consists of $n_1$ $V_i^{1,\alpha}$ and $n_2$ $V_i^{2,\alpha}$ fingerprints, namely $V_i^\alpha = \{V_i^{1,\alpha,(1)}, V_i^{1,\alpha,(2)}, \ldots V_i^{1,\alpha,(n1)}, V_i^{2,\alpha,(1)}, V_i^{2,\alpha,(2)}, \ldots V_i^{2,\alpha,(n2)}\}$. To extend such fingerprint set to single-component systems, one could follow a similar approach as above, whereby the fingerprints contains components for each atom type. For the binary system like SiO2, the fingerprint set of Si can be represented as $V_i^\alpha = \{V_i^{1,\alpha,(1,Si-Si)}, \ldots V_i^{1,\alpha,(n1, Si-Si)}, V_i^{1,\alpha,(1,Si-O)}, \ldots V_i^{1,\alpha,(n2, Si-O)}, V_i^{2,\alpha,(1,Si-SiSi)}, \ldots V_i^{2,\alpha,(n3, Si-SiSi)}, V_i^{2,\alpha,(1,Si-SiO)}, \ldots V_i^{2,\alpha,(n4, Si-SiO)}, V_i^{2,\alpha,(1,Si-OO)}, \ldots V_i^{2,\alpha,(n5, Si-OO)}\}$. Similarly, the fingerprint set of O is represented as $V_i^\alpha = \{V_i^{1,\alpha,(1,O-Si)}, \ldots V_i^{1,\alpha,(n1, O-Si)}, V_i^{1,\alpha,(1,O-O)}, \ldots V_i^{1,\alpha,(n2, O-O)}, V_i^{2,\alpha,(1,O-SiSi)}, \ldots V_i^{2,\alpha,(n3, O-SiSi)}, V_i^{2,\alpha,(1,O-SiO)}, \ldots V_i^{2,\alpha,(n4, O-SiO)}, V_i^{2,\alpha,(1,O-OO)}, \ldots V_i^{2,\alpha,(n5, O-OO)}\}$.

### C. Machine learning techniques

Kernel regression and linear regression have been used in the construction of ML force fields.[6,12] In this work, we tried to construct (i) linear model (LM) and (ii) neural network model (NNM).

Due to the simplicity and speed, LM was developed to describe the linear dependence between the structural fingerprints and forces. The linear model takes the form of:

$$F_i^\alpha = w_1 V_i^{\alpha,(1)} + w_2 V_i^{\alpha,(2)} + \ldots\ldots + w_n V_i^{\alpha,(n)} \quad (5)$$

$w_i$ is the regression coefficient. Regression coefficients are generally determined quickly using a standard least-squares technique. When matrix $V$ is including the fingerprints of the reference atomic environments and $F$ denotes the atomic forces obtained by DFT, the residual sum of squares $\|Vw\text{-}F\|_2^2$ is minimized in the linear regression ($\|\cdot\|_2$ denotes the $L_2$ norm).

The multi-layers NN is a mathematical model that mimics the structure and function of a biological neural network. It consists of a number of computational units, i.e. nodes, organized in several layers. The number of layers and the number of neurons per layer, determines the complexity of NN. The nodes in each layer is connected to the nodes in the neighboring layers via a connection coefficient, which is called "weights". For example, the complete functional form of a NN that contains three layers can be expressed as:

$$F_i^\alpha = f_1^2\left(b_1^2 + \sum_k w_{k1}^{12} \cdot f_k^1\left(b_k^1 + \sum_j V_i^{\alpha(j)} \cdot w_{jk}^{01}\right)\right) \quad (6)$$

where $w_{jk}^{01}$ is used for the weight connecting node $j$ in layer 0 with node $k$ in layer 1, $b_k^1$ and $f_k^1$ are the bias weight and the activation function on the node $k$ in layer 1, respectively. There are many possible choices for the activation functions such as sigmoid, hyper-tangent and linear. The NN architecture can be presented with a short notation specifying the nodes per layer, such as 10-5-5-5-1. The weights and bias of NN can be determined with gradient-based descent algorithms (such as conjugate gradient and Broyden–Fletcher–Goldfarb–Shanno[51]) using the root mean squared error ($\delta_{RMS}$) of force as the objective function.

### D. Selection of structural fingerprints

The accuracy and speed of ML force field, as well as other data-driven atomic simulation approaches, depends on the choice of structural descriptors. In principle, we can achieve higher accuracy by using more structural fingerprints, since they provides a sufficient structural distinction of inequivalent atomic environments to avoid fitting contradictory data.[13] However, the utilization of overset of the fingerprints leads to increase the computational complexity, and make the fitting more difficult. Therefore, it is very important to establish an automatic protocol to identify a set of fingerprints, which provides the best balance between computational cost and accuracy of predictions.

In this work, we tried to determine the parameters of fingerprints of linear model by selecting the important ones from a large pool of candidates. The selection is performed with LASSO[55] and GA. In this paper, we only discussed the fingerprints selection for the linear model because of its simplicity and fast training speed. However, we found that the selected fingerprint set also works very well for the neural network model. The discussion will be given in detail in section III and IV.

For preparation, a large set of fingerprints as candidates are created by spanning over all the meaningful sets of parameters. For example, we choose a several values of $\eta$ on a logarithmic scale, and several values of $R_s$ equidistantly from 0 to $R_c$. Then, the parameters of radial fingerprint are all possible combination of these $\eta$ and $R_s$ values. The candidate pools are given in the supplementary materials.

LASSO technique is frequently used in the selection of structural descriptors for the ML potentials.[5,56] The LASSO provides not only a linear relationship between the inputs and targets, but also a sparse representation with a small number of non-zero regression coefficients. The nonzero regression coefficients can be explained as the automatically-selected basis functions that contributes more essentially to the target properties. LASSO minimizes a penalized residual sum of squares expressed as $||Vw-F||_2^2 + \lambda ||w||_1$, where $||\cdot||_1$ denotes the $L_1$ norm. The parameter $\lambda$ controls the trade-off relationship between sparsity and accuracy. In this study, a variety of fingerprint sets were selected from candidates by changing $\lambda$ scan from 0.001 to 0.1 in a logarithmic grid.

Genetic algorithms are also widely used to solutions search problems by bio-inspired operators such as mutation, crossover and selection. In this research, we regard a set of $N$ fingerprints as an individual, then the selection of optimum individual can be done, based on the fundamental assumption that the force prediction error ($\delta_{RMS}$) is negatively correlated with its fitness. The details of GA algorithm are given in the Supplementary materials. It is worth noting that GA is easily trapped with the local minima, and usually it cannot get the global optimum solution. However, we can still obtain the accurate enough fingerprint sets through many times of trials.

By controlling the factor $\lambda$ of LASSO and $N$ of GA, we can equally obtain the dependence of the number of fingerprints $N$ on the force prediction error ($\delta_{RMS}$).

Generally, the $\delta_{RMS}$ would be gradually reduced and finally approach to the $\delta_{RMS}$ of linear model using all candidates. To obtain a set of fingerprints with good balance between computational cost and accuracy, the selection criteria is set as the smallest fingerprint set whose force prediction error is not larger than 105% of $\delta_{RMS}$ of the linear model using all candidates.

**E. Linear mixture model**

In this study, the idea of mixture model, which recently proposed in the construction of energy-based ML potentials is used to establish ML force field.[44] Firstly, the atomic fingerprint vector $V_i^\alpha$ was assigned to $m$ clusters with Gaussian mixture model (GMM). Different from the $k$-means clustering which assign the atomic environment to specific group, the GMM provides the probability of vector $V_i^\alpha$ belongs to group $m$, which is expressed as:

$$p(V_i^\alpha | m) = \frac{a_m \Phi(\vec{V}_i^\alpha; \mu_m, \Sigma_m)}{\sum_m a_m \Phi(\vec{V}_i^\alpha; \mu_m, \Sigma_m)} \quad (7)$$

where $\Phi$ is a multivariate Gaussian distribution with mean $\mu_m$ and covariance matrix $\Sigma_m$, and the weight coefficient $a_m$ satisfying $\Sigma_m a_m = 1$. The $\mu_m$, $\Sigma_m$ and $a_m$ are determined by the expectation-maximization(EM) algorithm. Based on the result of GMM, an atomic environment can be assigned to the group with highest probability, and a distinct linear model was built to represent the dependence between the fingerprints and forces in each group. Finally, the atomic force can be expressed as:

$$F_i^\alpha = \sum_m F_i^{\alpha,m} \cdot p(V_i^\alpha | m) \quad (8)$$

where $F_i^{\alpha,m}$ is the atomic force calculated with linear regression model of group $m$:

$$F_i^{\alpha,m} = w_1^m V_i^{\alpha,(1)} + w_2^m V_i^{\alpha,(2)} + \ldots\ldots + w_n^m V_i^{\alpha,(n)} \quad (9)$$

Instead of exclusively using the reference data assign to a group, the linear model of each group was trained with all the reference data with weighted least squared techniques, and the probability $p(V_i^\alpha|m)$ was regard as weight. The number of groups was determined following certain criteria, such as Akaike information criterion (AIC)[57] and the Bayesian information criterion (BIC)[58].

## III. Cu MACHINE LEARNING FORCE FIELD

### A. Linear Model

The cutoff distance $R_c$ of fingerprints of Cu atomic environment is determined to be 6.5 Å according to a convergence test. The detailed information of cutoff distance convergence test can be seen in the supplementary material in Fig. S1.

To demonstrate that angular fingerprints (eq. 4) makes the accuracy of force prediction improve, we generate two sets of fingerprints: set 1 and 2. The set 1 has only radial fingerprints (eq. 3), while the set 2 has both radial and angular fingerprints (eq. 3 and 4). Both set 1 and 2 contain 16 fingerprints with different parameters. The parameters are determined empirically based on the principle that the functional shape form a fine and even grid around the center atom (see supplementary materials for the parameters and functional curve of set 1 and 2). The linear models (LM) are fitted based on set 1 and 2, and the force prediction errors $\delta_{RMS}$ are shown in Fig. 3. The $\delta_{RMS}$ and $\delta_{MAE}$ of the linear model base on set 1 is 0.095 eV/Å and 0.045 eV/Å, respectively, which are comparable with those of other contemporary ML force field for elemental materials.[12,13,59] But the $\delta_{RMS}$ of LM with set 2 is much smaller, even though the set 2 contain the same number of fingerprints with set 1. The $\delta_{RMS}$ and $\delta_{MAE}$ of set 2 is 0.051 eV/Å and 0.027 eV/Å, respectively. We can clearly see that the use of the angular fingerprint makes the force prediction much more accurate.

In the next step, we carried out the selection of fingerprints with LASSO and GA following the protocol we discussed in section II. We generate two large pools of fingerprints, *i.e.* set 3 and 4. The set 3 has 119 radial functions, and the set 4 has 117 radial and angular functions. The $\delta_{RMS}$ of LM with sets 3 and 4 is 0.086 and 0.039 eV/Å, respectively. Not surprisingly, the LMs with set 3 and 4 give better performance than that with set 1 and 2, since one can describe the atomic environment more and more complete by increasing the number of fingerprints, with the expenses of larger computational costs. To identify a small subset that conveys the essential structural information, the LASSO and GA methods are used to select fingerprints from the set 3 and 4. The dependence of $\delta_{RMS}$ on the number of fingerprints N is plotted in the Fig. 4. As shown in Fig 4(a), the $\delta_{RMS}$ of a set of fingerprints that contains 15(16) LASSO(GA) selected ones is 0.088 (0.087) eV/Å. Thus, the both selected sets perform better than the manually selected set 1. Similarly, the GA selected fingerprint set from set 4 also performs better than manually selected set 2, with the same number of fingerprints. The GA automatic selection protocol is more effective than the frequently used LASSO, since the GA selected set always has lower $\delta_{RMS}$ than the LASSO selected one when the number of fingerprints in the same. But the computational cost of GA protocol is higher than LASSO, since it involves many times of LM training and prediction. It is also obvious that the selection curve of set 4 converges slower than set 3, which implies that set 4 contains more complicate structural information. By setting the threshold $\delta_{RMS}$ as 105% of the $\delta_{RMS}$ of the set 3 and 4, we can select 15 and 8 fingerprints from set 3, 86 and 32 fingerprints from set 4 with LASSO and GA respectively. These selected fingerprint sets are named as set 5 – 8.

## B. Neural Network Model

Based on the fingerprint set 1 – 8, ML force fields are constructed by neural networks (NN). As we know, NN can have different functional forms when using different topology and activation functions. Here, two types of NN model (NNM) are used. The first one, $NNM_1$, uses linear activation functions in all its nodes and do not use bias for nodes. The second type, $NNM_2$, uses hyper-tangent activation function in hidden nodes, linear activation function in output nodes. All the NN topology have 3 hidden layers and 20 nodes in each layer, which results in 290 – 1431 weight parameters. The $\delta_{RMS}$ of LM, $NNM_1$ and $NNM_2$ are summarized in the Table I.

We are surprised to find that even though more complicate functional form is used, the $\delta_{RMS}$ of NNM is only very slightly reduced from the LM. It is contrary to the popular belief that the high order, like NN, regressor usually perform substantially better than the low order regressor, like LM, in many ML models of material properties.[60] Maybe, the LM can predict force well because the force is a not very complex function in terms of the atomic positions. Another successful construction of ML force field with linear regression can be seen in Ref [6]. The $NNM_2$ force field with set 4 is the most accurate, and its $\delta_{RMS}$ is 0.40 eV/Å in the training set and 0.038 eV/Å in the testing set. However, the LM with set 8 has very close accuracy, while its computational cost in regression process and force evaluation is much lower than the previous one. In addition, we noticed that the NNMs with set 5 and 6 have higher force prediction accuracy than that with set 1. It implied that the GA and LASSO selected fingerprint set also works for the NN models even though they are selected according to the $\delta_{RMS}$ of LM.

It is noteworthy that the force fields follow certain restrictions in the rotation operation. If a rotation operation is acting on a structure, the predicted force must transform accordingly. To ensure the ML force field abide by this rule, several solutions has been used. A. Glielmohe *et al.* proposed a new family of vector kernels of covariant nature which can obey the correct vector behavior of forces under symmetry transformations.[46] The solution of Botu *et al.* is more straightforward. They added new training data by rotating the collected atomic configurations and forces.[11] However, we found LM and $NNM_1$ can instinctively follow this rule, because both the fingerprint and the functional forms transform in the identical manner upon rotation. The $NNM_1$ is essentially a linear regression method since the nonlinear activation function is not used in any layer of nodes. The discussion is given in the supplementary material.

## C. Linear Mixture Model

The ML force field for Cu that base on mixture model method is constructed with fingerprint set 8. The fingerprints of all the atomic environments in the reference database are assigned to 9 groups with certain probability using Gaussian mixture model method. The number of samples in each individual group varies from 50415 to 771684. We don't need to worry about the risk the overfitting because the number of training data is so much more than the number of fitting parameters even in the smallest group. We found that the Gaussian mixture model clustering the atomic environments with clear physical meanings. The detailed information is given in the supplementary material. Figure 5 illustrates the comparison of the DFT calculated atomic forces with the forces predicted with the linear mixture model (LMM). The $\delta_{RMS}$ is 0.035 eV/Å for both training and testing sets, and coefficient of determination is 0.9992.

The LMM surpass LM and NNM for two reasons. First, the reference database contains highly diverse reference structures, which makes the fitting by a single model difficult. The LMM separates the whole training set into different groups, and accordingly, the structure diversity inside one group can be effectively reduced. Second, the fingerprints are very unevenly distributed in the configuration space. In this case, ML techniques, such as LM, gives more weight to the configuration space where the training data is densely distributed. Actually, Huan *et al.* proposed a training data selection algorithm based on K-means clustering to ensure uniform and diverse sampling from all regions of the configuration space.[13] However, we found the mixture model method fundamentally resolve this problem because it constructs the separated ML models for different region of configuration space, and the difference of sample density in the configuration space would not affect the fitting of ML models.

## IV. SiO$_2$ MACHINE LEARNING FORCE FIELD

### A. Linear model

Similar with the process of constructing Cu ML force fields, we constructed the LM for binary material SiO$_2$. The cutoff distance $R_c$ of fingerprints was determined to be 8.0 Å according to the convergence test (see supplemental material for details).

Similar to the force field construction for Cu, we constructed two sets of fingerprints with similar number of fingerprints: set 1, 30 radial fingerprints; set 2, 37 radial and angular fingerprints. The parameters of each set of fingerprints are also listed in the supplementary material. The $\delta_{RMS}$ of LM with set 1 and 2 are shown in Fig. 6. As expected, the set 2 has better performance than set 1, which further demonstrated the importance of angular fingerprints in improve force prediction accuracy. It is also obvious that that the existence of angular fingerprints has greater impact on the force prediction of Si atoms. It may be because one Si is bonded with four O with covalent bonds and form a tightly bonded tetrahedron unit. So, the force of Si atoms has great dependence on the bond.

As we proved with Cu, the GA fingerprints selection protocol outperform LASSO in most case. Here, the GA selection is carried out using the set 4 (2112 radial and angular fingerprints) as initial pool of fingerprints. Figure 7 shows that it is possible to progressively improve the force prediction of both O and Si atoms by making the selection more inclusive. Using the same criterion (105% of $\delta_{RMS}$ of set 4), we determined the set 5, which contains 256 fingerprints for both O and Si. The $\delta_{RMS}$ of set 5 is 0.219 eV/Å in the training set and 0.219 eV/Å in the testing set, while the mean absolute error $\delta_{MAE, F}$ is 0.120 eV/Å in the training set and 0.122 eV/Å in the testing set. Such value is slightly larger than the other ML force fields for binary materials (e.g. a-SiO$_2$ $\delta_{MAE}$ 0.06 eV/Å;[8] a-Al$_2$O$_3$ $\delta_{MAE}$ 0.05 eV/Å;[12] and monoclinic - HfO$_2$ $\delta_{MAE}$ 0.05 eV/Å[12]), because the structural diversity of our reference database is significantly higher than the previous studies (if we choose only pristine, single-phase reference data, the $\delta_{MAE}$ can be reduced to 0.04 – 0.08 eV/Å). Similarly, the GA selected fingerprint sets have better performance than the manually determined ones, *i.e.* set 2.

### B. Neural network and linear mixture model

The neural network models (both NNM$_1$ and NNM$_2$) were established using set 1-5, and the results obtained for $\delta_{RMS}$ are shown in the Table II. The NN configurations have 3 hidden layers and 5 nodes in each layer. Since the set 4 contains more than 2000 fingerprints, the corresponding NN will have too much weight parameters to fit. So, such case is not considered here. Similar with the Cu case, we cannot see obvious the improvement on $\delta_{RMS}$ by using NNM instead of LM.

The LMM was constructed for SiO$_2$ base on the fingerprint set 5. The fingerprints database is assigned to 21 groups using gaussian mixture model method, and the smallest group contains 5457 O (4271 Si) environment as training data. Figure 8 illustrates the comparison of the DFT-calculated atomic forces with the forces predicted with LMM. The $\delta_{RMS}$ is 0.120 eV/Å (O, 0.106; Si, 0.149) for both training set and 0.124 eV/Å (O, 0.108; Si, 0,156) for the testing set, and the $\delta_{MAE,F}$ is 0.081 eV/Å for training set and 0.084 eV/Å for testing set.

It is worth noting that the Coulomb interaction plays important role in the multicomponent materials due to the significant charge transfer among difference elements. But such interaction is not fully considered with the current ML force field, because we truncate the interatomic interaction within the cutoff radius. To tickle this problem, we made preliminary attempt to construct the ML force field that explicitly consider the Coulomb interaction by adding a second set of ML model to predict the environment-dependent charges at respective atomic sites.

Then the short-range force (the contribution inside of cutoff sphere) and long-range force (the contribution outside of cutoff sphere) are calculated respectively. The details and results are provided in the supplementary materials. Surprisingly, we found that charge-explicit ML force field does not achieve a better accuracy than the ordinary one. Actually, it is consistent with Artrith's work on the zinc oxide NN potential, in which the charge prediction and Coulomb interaction calculation do not makes energy prediction better.[61] In addition, we note that the long-range part is actually calculated with precisely, and the error mainly comes from the prediction of short-range part.

### V. Application of ML force fields

To prove the fidelity of ML force fields in the atomic simulation, various atomic simulations, including structural optimization, molecular dynamics and nudge elastic band calculation, was carried out, and many derived properties were compared with *ab initio* calculations in the same condition. The LM, $NNM_2$ and LMM for Cu and $SiO_2$ are all used in the atomic simulations.

The molecular dynamics simulation started with the equilibrium configuration of *fcc*-Cu (256 atoms), quartz-$SiO_2$ (144 atoms) and cristobalite $SiO_2$ (192 atoms). NVT ensemble is performed at the temperature of 1000 K. The MD trajectories were accumulated for 5 ps using a time step of 1 fs. To directly assess the performance of our ML force fields in MD, the radial distribution function (RDF) and angular distribution function (ADF) averaged over the last 2.5 ps. The results are shown in Fig. 9 and 10 and compare with the *ab initio* MD. The agreement for both is excellent. It should be noted that our ML force fields are trained with the configurations with much smaller size, which demonstrates the transferability of ML force fields to the large system.

Next, we discuss the applicability of the ML force fields in geometry optimization. The structural optimization starts from the non-equilibrium pristine and defective *fcc*-Cu, quartz- and cristobalite- $SiO_2$, obtained from 300K MD. The steepest descent method is adopted to relax the nuclei positions until the maximum force acting on one atom is less than 0.001 eV/Å. We found the atoms in pristine non-equilibrium geometry recover to the correct symmetry positions. After that, the final structures are compared with the DFT-optimized ones with the same force tolerance criterion, and the maximum and average nuclei position discrepancy between DFT- and ML force field-optimized structures are listed in Table III. The ML force fields show excellent accuracy in the optimization of pristine crystals, and the maximum discrepancy is less than 0.001 Å. For the defective structures, the position discrepancy is a little bigger, but maximum difference is still less than 0.06 Å.

Another atomic simulation we tested is nudged elastic band method. Here, the energy profile along the vacancy migration pathway in the bulk Cu is computed and the results are shown in Fig. 11. The migration activation energy is estimated to be 0.74 eV, 0.74 eV and 0.73 eV with LM, NNM and LMM respectively. Both values are very closed to the DFT result, 0.72 eV. Due to the very high barrier energy of vacancy and interstitial diffusion, the migration transient configurations are rarely sampled in the process of reference data gathering. In this case, the ML force fields constructed in this work might cannot be reliable in the NEB calculation. The special reference data sampling method, which discussed in our previous work,[30] can solve this problem.

### VII. CONCLUSIONS

In this work, the ML force fields were constructed for both elemental (Cu) and binary ($SiO_2$) materials. Firstly, the atomic environment is represented with the structural fingerprints considering both interatomic distance and bond angles. Then we discuss automatic protocols to select a number of fingerprints out of a large pool of candidates, based on LASSO and GA. After that, the ML force fields were constructed with difference ML techniques, including linear regression, neural network, and recently-developed mixture model method. Lastly, the ML force fields were compared with the DFT in structural optimization, MD simulation and nudged elastic band calculation.

We found that the new structural fingerprints can significantly improve the accuracy of ML force fields for both Cu and $SiO_2$. The ML force fields based on the GA-selected fingerprints outperform that based on empirically selected fingerprint set with same size. In addition, the force fields built with mixture model method has the highest force prediction accuracy among the different ML techniques. The force fields of Cu ($SiO_2$) constructed with mixture model method has $\delta_{RMS}$ of 0.035 (0.124) eV/Å. At last, we demonstrated the fidelity of ML force fields in some atomistic simulations, such as geometry optimization of atomic structures, molecular dynamics and nudged elastic band calculation to determine vacancy diffusion barriers.

### SUPPLEMENTAL MATERIAL

See supplementary material for the parameters of structural fingerprints, genetic algorithm automatic selection protocol, ML force fields with consideration of Coulomb interaction, local structure of different GMM groups, composition of reference database and comparison of ML force field with the Behler-Parrinello neural network potential.

**ACKNOWLEDGEMENTS**

This paper is based on results from a project (P16010) commissioned by the New Energy and Industrial Technology Development Organization (NEDO).

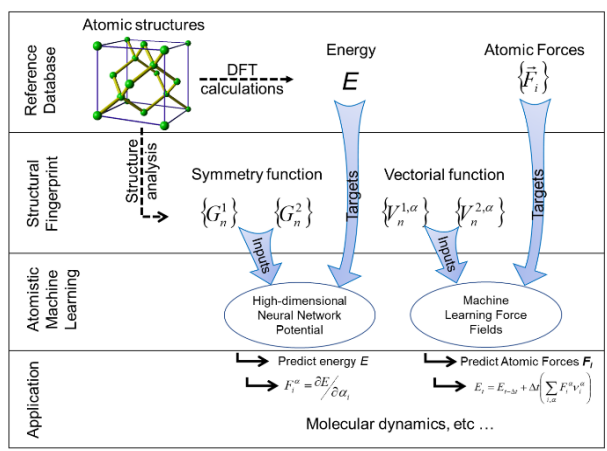

FIG. 1. The scheme of machine learning force field and high-dimensional neural network potential.

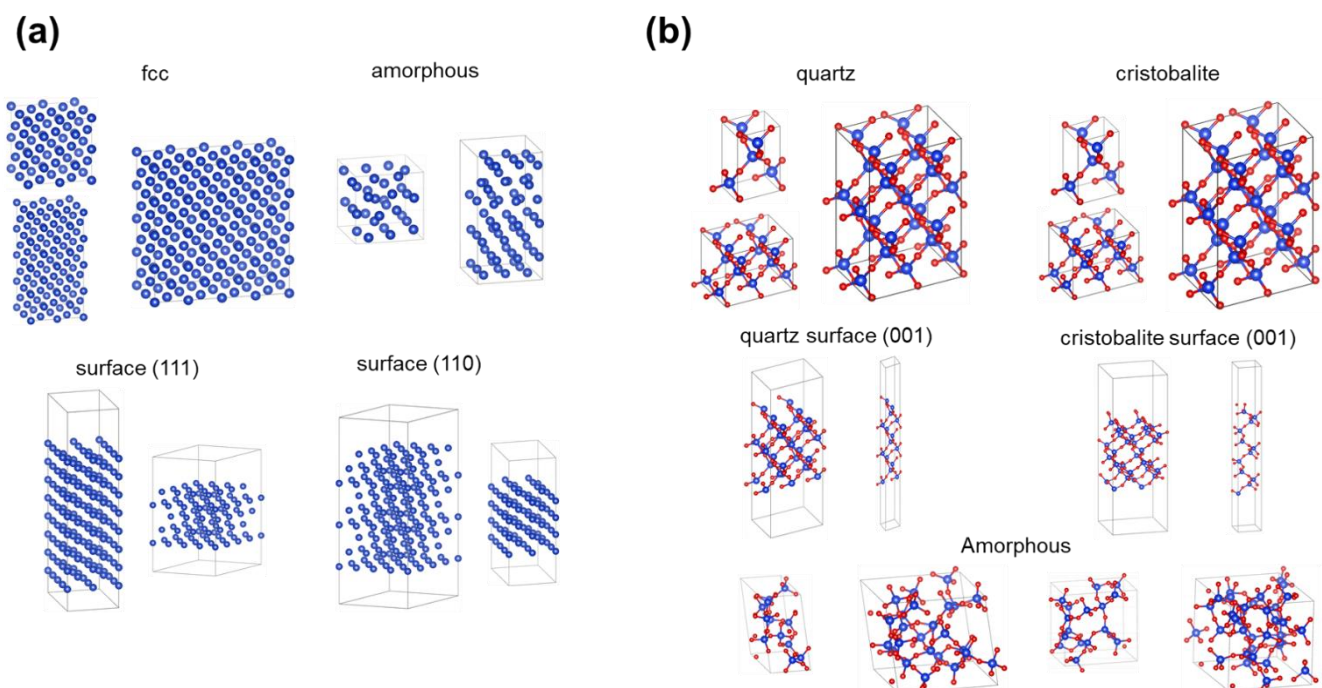

FIG. 2. Reference configurations used to sample atomic environments for training and testing of machine learning force fields; (a) Cu, (b) $SiO_2$.

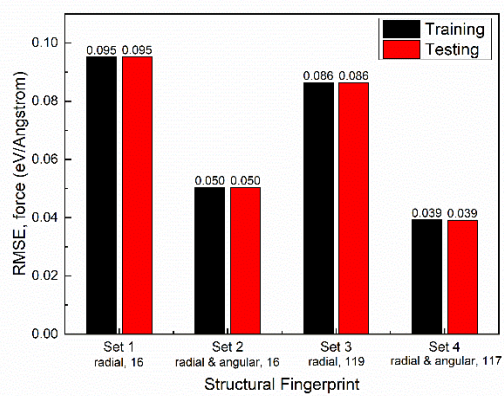

FIG. 3. The RMSEs of force in training and testing sets obtained for the Cu linear regression force fields that constructed with structural fingerprint set 1 – 4.

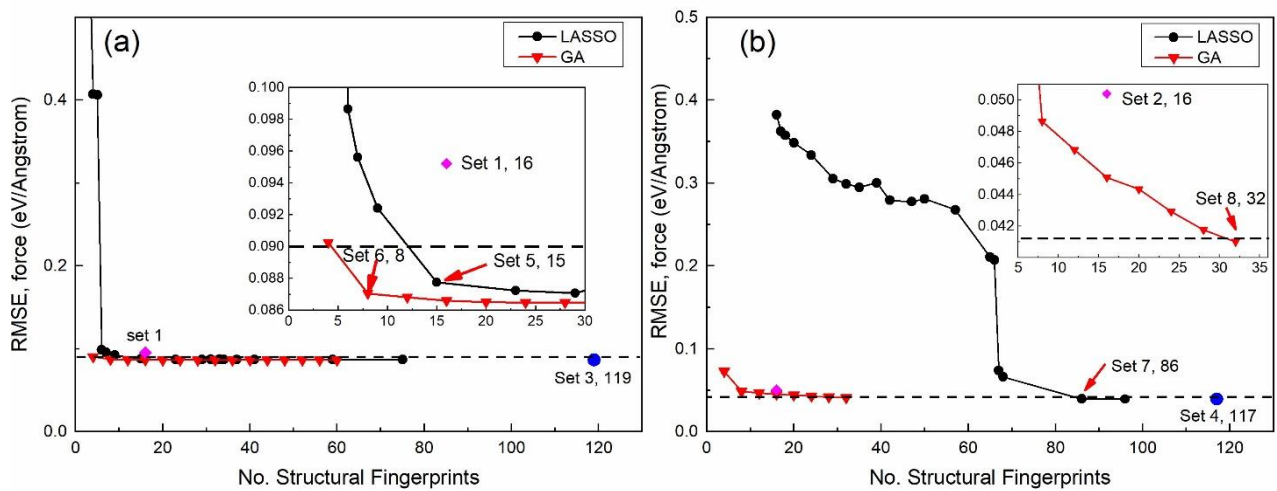

FIG. 4. Dependence of RMSE of force of the Cu linear regression force fields on the number of LASSO- or GA- selected fingerprints. (a) the fingerprints are selected from 119 radial fingerprints of set (3); (b) the fingerprints are selected from 117 radial and angular fingerprints of set (4).

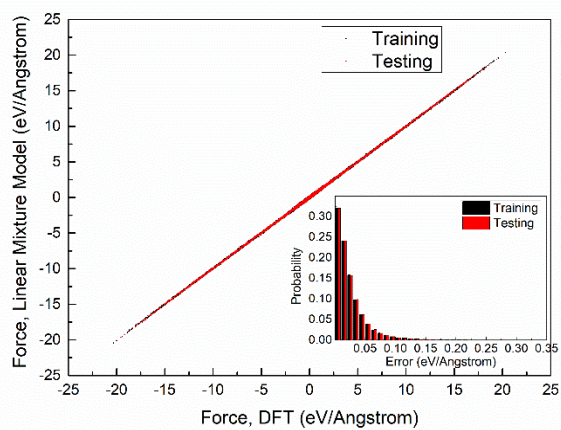

FIG. 5. Comparison of the forces predicted using the Cu mixture model ML force field with reference DFT results. The insert shows the distribution of the prediction errors.

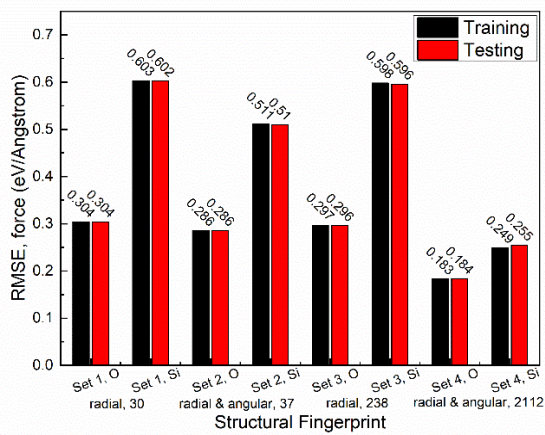

FIG. 6. The RMSEs of force prediction in training and testing sets obtained for the $SiO_2$ linear regression force fields that constructed with structural fingerprint set (1) – (4).

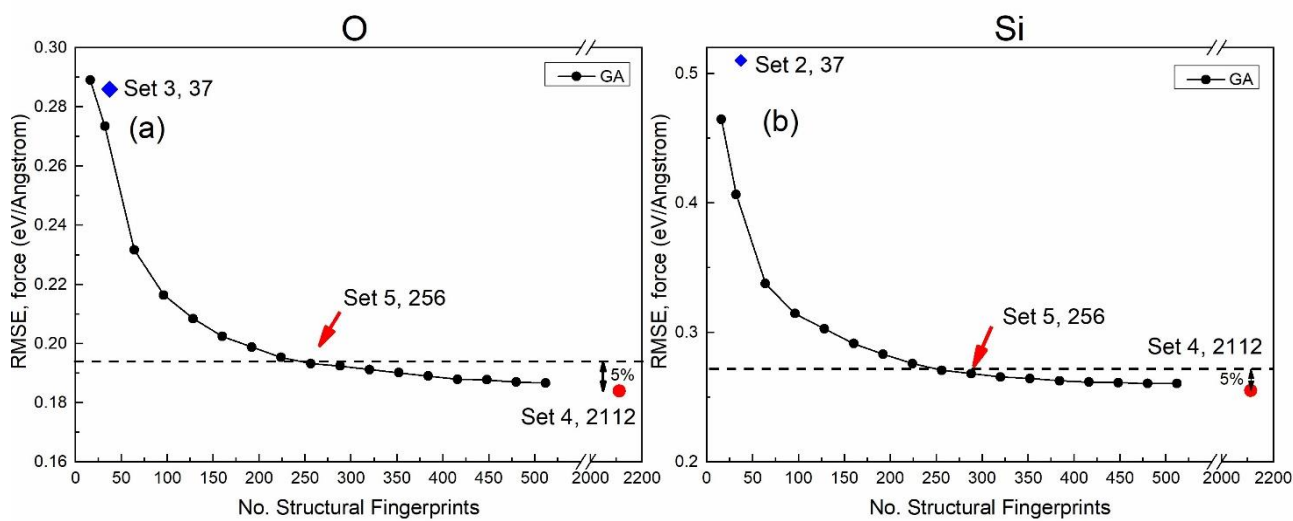

FIG. 7. Dependence of RMSE of force acting on O and Si atoms on the number of LASSO- or GA- selected fingerprints. (a) the fingerprints of O atomic environment are selected from set (4); (b) the fingerprints of Si atomic environment are selected from set (4).

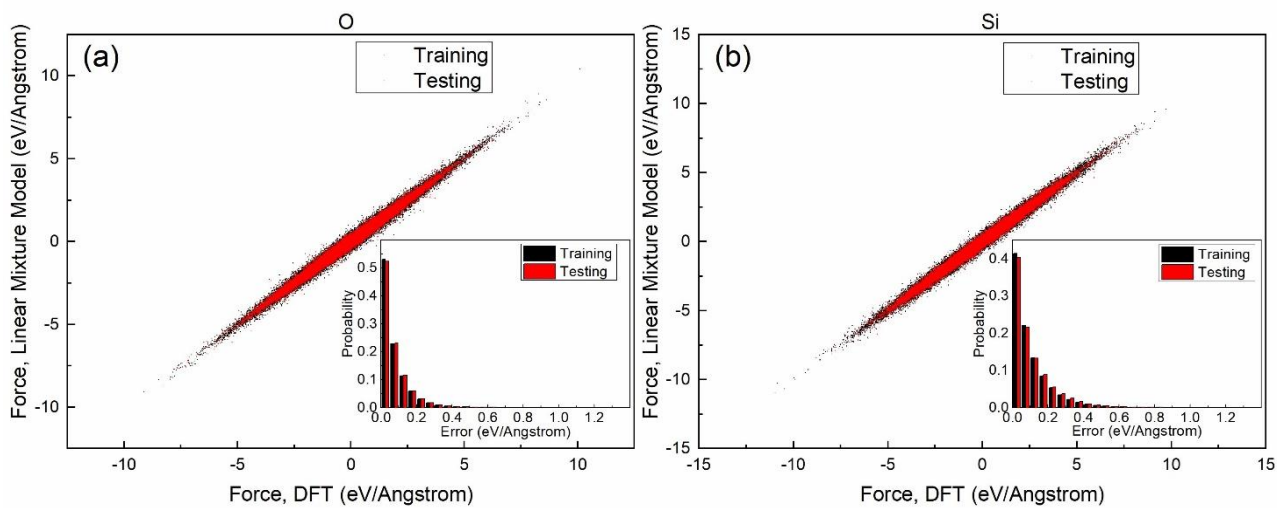

FIG. 8. Comparison of the forces predicted using the SiO$_2$ mixture model ML force field with reference DFT results. The inserts show the distribution of the prediction errors.

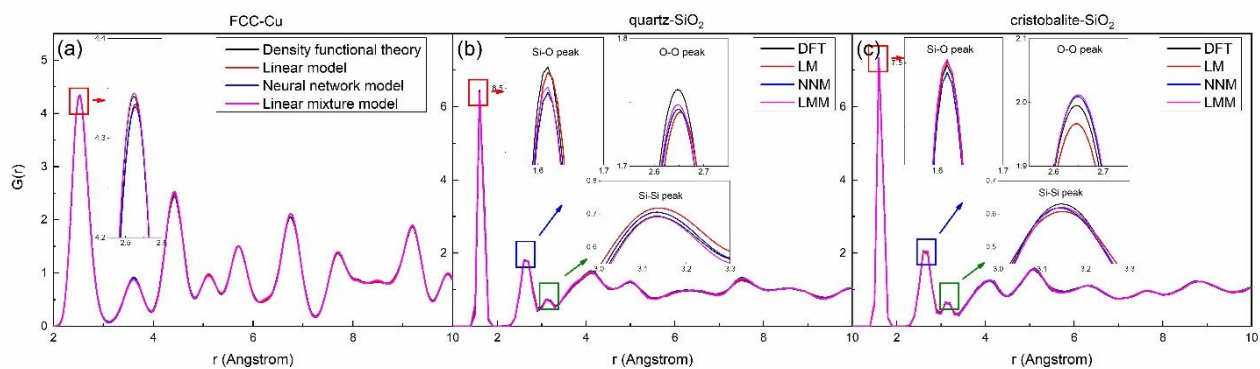

FIG. 9. The radial distribution function averaged over the 2.5 ps MD simulations based on density functional theory, linear regression / mixture model force fields and neural network potential. (a) fcc-Cu; (b) quartz-SiO$_2$; (c) cristobalite-SiO$_2$.

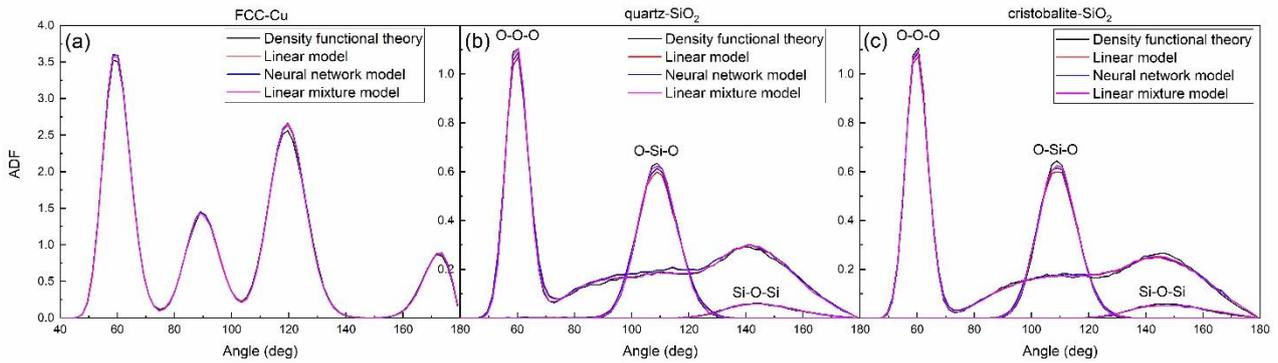

FIG. 10. The angular distribution function averaged over the 2.5 ps MD simulations based on density functional theory, linear regression / mixture model force fields and neural network potential. (a) fcc-Cu; (b) quartz-SiO$_2$; (c) cristobalite-SiO$_2$. The angles are determined with the maximum bond length of 3.0 Å.

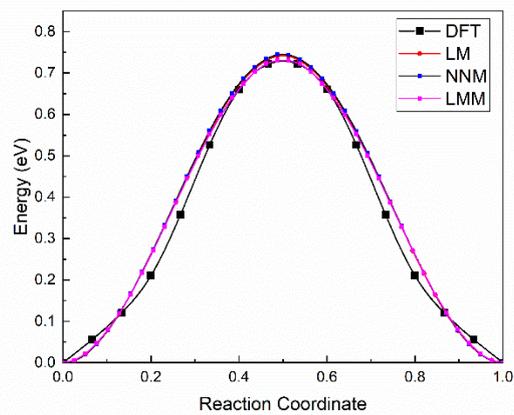

FIG. 11. The potential energy profile along the vacancy migration pathway in bulk Cu, which obtained in NEB calculation.

TABLE I. The RMSEs of force in training and testing sets obtained for the Cu ML force fields that constructed that constructed with different fingerprint sets and ML techniques.

| Fingerprint | Num. of Fingerprints | $\delta_{RMS}$ (eV/Å) | | | | | |
|---|---|---|---|---|---|---|---|
| | | Linear regression | | Neural Network 1 | | Neural Network 2 | |
| | | Training | Testing | Training | Testing | Training | Testing |
| Set 1 | 16 | 0.095 | 0.095 | 0.092 | 0.091 | 0.092 | 0.092 |
| Set 2 | 16 | 0.050 | 0.050 | 0.049 | 0.048 | 0.048 | 0.050 |
| Set 3 | 119 | 0.086 | 0.086 | 0.085 | 0.086 | 0.085 | 0.085 |
| Set 4 | 117 | 0.039 | 0.039 | 0.039 | 0.041 | **0.040** | **0.038** |
| Set 5 | 15 | 0.088 | 0.088 | 0.086 | 0.086 | 0.087 | 0.085 |
| Set 6 | 8 | 0.087 | 0.087 | 0.086 | 0.088 | 0.088 | 0.088 |
| Set 7 | 86 | 0.041 | 0.040 | 0.040 | 0.039 | 0.039 | 0.040 |
| Set 8 | 32 | **0.041** | **0.041** | 0.041 | 0.040 | 0.041 | 0.041 |

TABLE II. The RMSEs of force in training and testing sets obtained for the SiO$_2$ ML force fields that constructed that constructed with different fingerprint sets and ML techniques. The values in parentheses represent the RMSEs for O and Si respectively.

| Fingerprint | Num. of Fingerprints | $\delta_{RMS}$ [$\delta_{RMS,O}$/ $\delta_{RMS,Si}$ ] (eV/Å) | | | | | |
|---|---|---|---|---|---|---|---|
| | | Linear regression | | Neural Network 1 | | Neural Network 2 | |
| | | Training | Testing | Training | Testing | Training | Testing |
| Set 1 | 30 | 0.404 [0.304/0.603] | 0.403 [0.304/0.602] | 0.398 [0.299/0.596] | 0.397 [0.297/0.597] | 0.390 [0.291/0.589] | 0.393 [0.293/0.592] |
| Set 2 | 37 | 0.361 [0.286/0.511] | 0.361 [0.286/0.510] | 0.352 [0.278/0.502] | 0.351 [0.274/0.505] | 0.351 [0.276/0.501] | 0.350 [0.275/0.501] |
| Set 3 | 238 | 0.397 [0.297/0.598] | 0.396 [0.296/0.596] | 0.388 [0.287/0.589] | 0.394 [0.293/0.595] | 0.385 [0.284/0.587] | 0.395 [0.295/0.596] |
| Set 4 | 2112 | **0.204** [0.183/0.249] | **0.209** [0.184/0.255] | - | - | - | - |
| Set 5 | 256 | **0.219** [0.194/0.270] | **0.219** [0.194/0.271] | 0.216 [0.190/0.267] | 0.220 [0.195/0.269] | 0.214 [0.188/0.267] | 0.218 [0.192/0.270] |

TABLE III The maximum nuclei position disagreement between DFT-optimized and ML force field-optimized. The value in parentheses indicates the average disagreement.

| | | Linear regression (Å) | Neural network model (Å) | Mixture model (Å) |
|---|---|---|---|---|
| Cu fcc | pristine | 0.0004 [0.0002] | 0.0002 [0.0001] | 0.0002 [0.0001] |
| | vacancy | 0.0044 [0.0029] | 0.0047 [0.0031] | 0.0043 [0.0027] |
| | interstitial | 0.0017 [0.0016] | 0.0019 [0.0013] | 0.0015 [0.0012] |
| SiO2 quartz | pristine | 0.0057 [0.0052] | 0.0052 [0.0043] | 0.0049 [0.0041] |
| | O vacancy | 0.033 [0.016] | 0.035 [0.017] | 0.030 [0.015] |
| | O interstitial | 0.042 [0.018] | 0.044 [0.018] | 0.041 [0.016] |
| SiO2 cristobalite | pristine | 0.0069 [0.0061] | 0.0067 [0.0060] | 0.0057 [0.0050] |
| | O vacancy | 0.031 [0.013] | 0.034 [0.015] | 0.029 [0.012] |
| | O interstitial | 0.041 [0.013] | 0.043 [0.017] | 0.042 [0.014] |